\documentclass{PoS}

\def \inte {$INTEGRAL$}

\def \suz {$Suzaku$}

\def \src {IGR\,J16479--4514}
\def \ergsec{\hbox{erg s$^{-1}$}}
\def \ferg {erg cm$^{-2}$ s$^{-1}$}
\def \hcm {\hbox {\ifmmode $ atom cm$^{-2}\else atom cm$^{-2}$\fi}}

\def \msun {M$_{\odot}$}

\title{The Supergiant Fast X-ray Transient with the shortest orbital period:
Suzaku observes one orbit in IGR J16479-4514}

\ShortTitle{The SFXT with the shortest orbital
period: Suzaku observes one orbit in IGR J16479-4514}

\author{\speaker{Lara Sidoli} 
\\
        INAF, Istituto di Astrofisica Spaziale e Fisica Cosmica, \\
         Via E.\ Bassini 15,   I-20133 Milano,  Italy\\
        E-mail: \email{sidoli@iasf-milano.inaf.it}}

\author{P.~Esposito,$^a$ V.\ Sguera,$^b$ A.\ Bodaghee,$^c$ J.A.\ Tomsick,$^c$ K.\ Pottschmidt,$^{d,e}$ J.~Rodriguez,$^f$ 
P.\ Romano,$^g$ J.\ Wilms,$^h$   \\
\llap{$^a$}INAF-IASF Milano, Italy \\
\llap{$^b$}INAF-IASF Bologna, Italy\\
\llap{$^c$}SSL, University of California, Berkeley, CA, USA\\
\llap{$^d$}CRESST-NASA/GSFC, Greenbelt, MD, USA\\
\llap{$^e$}CSST, University of Maryland Baltimore County, Baltimore, MD, USA\\
\llap{$^f$}AIM - Univ. Paris VII and CEA Saclay, France \\
\llap{$^g$}INAF-IASF Palermo, Italy\\
\llap{$^h$}Remeis-Observatory Bamberg, Germany\\
}

\abstract{The eclipsing hard X--ray source \src\ is the  Supergiant Fast X-ray Transient (SFXT)
with the shortest orbital period (P$_{orb}$$\sim$3.32 days). 
This allowed us to perform a 250~ks long X--ray observation with  \suz\  in 2012 February,
covering most of its orbit, including the eclipse egress. 
Outside the eclipse, the source luminosity is around a few 10$^{34}$\ergsec. 
The X--ray spectrum can be fit with an absorbed power law together with
a neutral iron emission line at 6.4 keV. The column density, N$_{\rm H}$, is constant at
$\sim$10$^{23}$~cm$^{-2}$ outside the X--ray eclipse.  During the eclipse it is lower,
consistent with a scattering origin for the low X--ray emission during the eclipse by the supergiant companion.
The scattered X--ray emission during the X--ray eclipse 
is used to directly probe the density, $\rho$$_{\rm w}$, of the companion wind at the orbital separation, 
resulting in $\rho$$_{\rm w}$=7$\times$10$^{-14}$~g~cm$^{-3}$, which translates into a
ratio $\dot{M}_{w}/ v_{\infty}=7\times$10$^{-17}$~M$_{\odot}$/km of the wind mass loss rate to
the wind terminal velocity. This ratio, assuming reasonable 
terminal velocities in the range 500--3000~km~s$^{-1}$, translates into an accretion luminosity 
two orders of magnitude
higher than that observed. 
We conclude that a mechanism reducing the accretion rate onto the compact object is at work, likely 
due to the neutron star magnetosphere.
}

\FullConference{An INTEGRAL view of the high-energy sky (the first 10 years) -
9th INTEGRAL Workshop and celebration of the 10th anniversary of the launch\\
                 15-19 October 2012\\
                 Bibliotheque Nationale de France, Paris, France}

\begin{document}

\section{$Suzaku$ observes \src}
 
\src\ was discovered   by \inte\ in 2003 (Molkov et al. 2003)
at hard energies (18--50 keV), as a source displaying
recurrent, transient and flaring activity reaching 600 mCrab (20--60 keV; Sguera et al. 2005, 2006).
The counterpart, accurately localized with $Chandra$ (Ratti et al. 2010),
is an O-type supergiant at a distance of 2.8$_{-1.7} ^{+4.9}$~kpc (Nespoli et al. 2008).
The peculiar X--ray properties together with the supergiant companion
suggested to classify the source as a member of the Supergiant Fast X--ray Transients (SFXTs).
\src\ shows X--ray eclipses  (Bozzo et al. 2008a) and is the SFXT which displays the shortest 
orbital period  (3.3193$\pm{0.0005}$\,days; Jain et al. 2009, Romano et al. 2009).
$Swift$XRT monitoring has revealed that its more usual luminosity state 
is at a level of  10$^{33}$-10$^{34}$~erg~s$^{-1}$ (1--10 keV), 
2--3 orders of magnitude lower than the  peak of the flares (Sidoli et al. 2008).

\begin{figure}[ht!]
\begin{center}
\vspace{0.truecm}
\includegraphics*[angle=-90,scale=0.65]{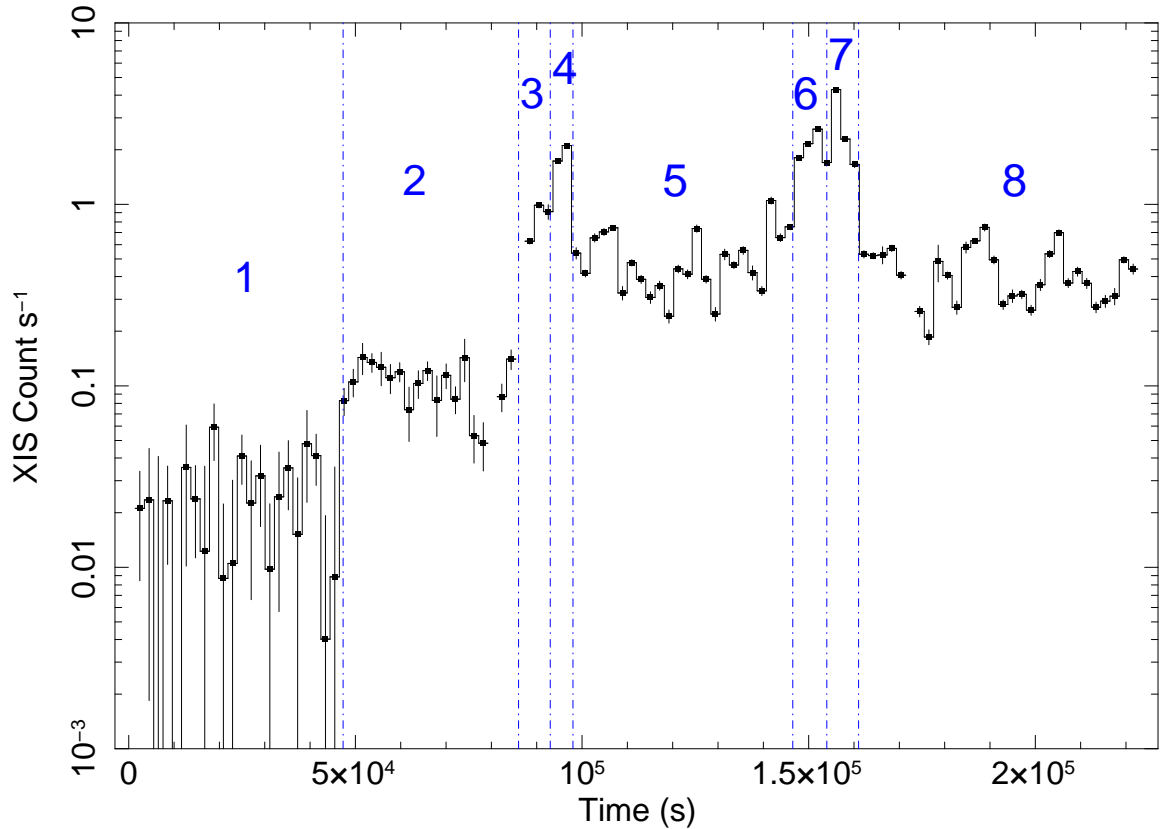}
\end{center}
\vspace{-0.truecm}
\caption{\scriptsize $Suzaku$/XIS (1--10 keV)  light curve of IGR~J16479-4514.  
Vertical lines and numbers indicate the time selected spectra. 
}
\label{lsfig:suz_lc}
\end{figure}

We observed \src\ with \suz\ between 2012 February 23 and 26 (Obs ID 406078010), 
at the X--ray Imaging Spectrometer (XIS; Koyama et al. 2007) nominal position for a net
exposure of 250 ks.
The main aim of the observation was to  obtain an in-depth investigation
of the   variability of the X--ray and supergiant wind properties
along the orbit (more details about the data reduction and analysis can
be found in Sidoli et al. 2013).

The 1--10 keV light curve  is displayed in Fig.~\ref{lsfig:suz_lc} and covers 
about 80\% of the orbit. Also the eight time intervals used to 
perform a temporal selected spectroscopy are marked with vertical lines.
The very low intensity state at the start of the observation 
is produced by the eclipse of the central X--ray source by the O-type supergiant. 
Two faint flares, lasting 10--15~ks, are present outside the eclipse,
reaching a peak flux of 3-4$\times10^{-11}$~\ferg, spaced by $\Delta \phi$=0.2 in orbital phase.
Interestingly, the first flare is located at a similar orbital phase as  other 
bright flares observed in the past from the same source (Bozzo et al. 2009).
This might suggest that   a phase-locked gas structure originating
from the supergiant is present along the orbit, triggering the enhanced accretion rate 
that produces the X--ray flare, as the compact object crosses it.
Alternatively, the orbit could be mildly eccentric (although the very short orbital period
favors a circular orbit) and the flare is triggered near periastron.

\begin{figure}[ht!]
\begin{center}
\vspace{0.7truecm}
\hspace{-1.0truecm}
\includegraphics*[angle=-270,scale=0.68]{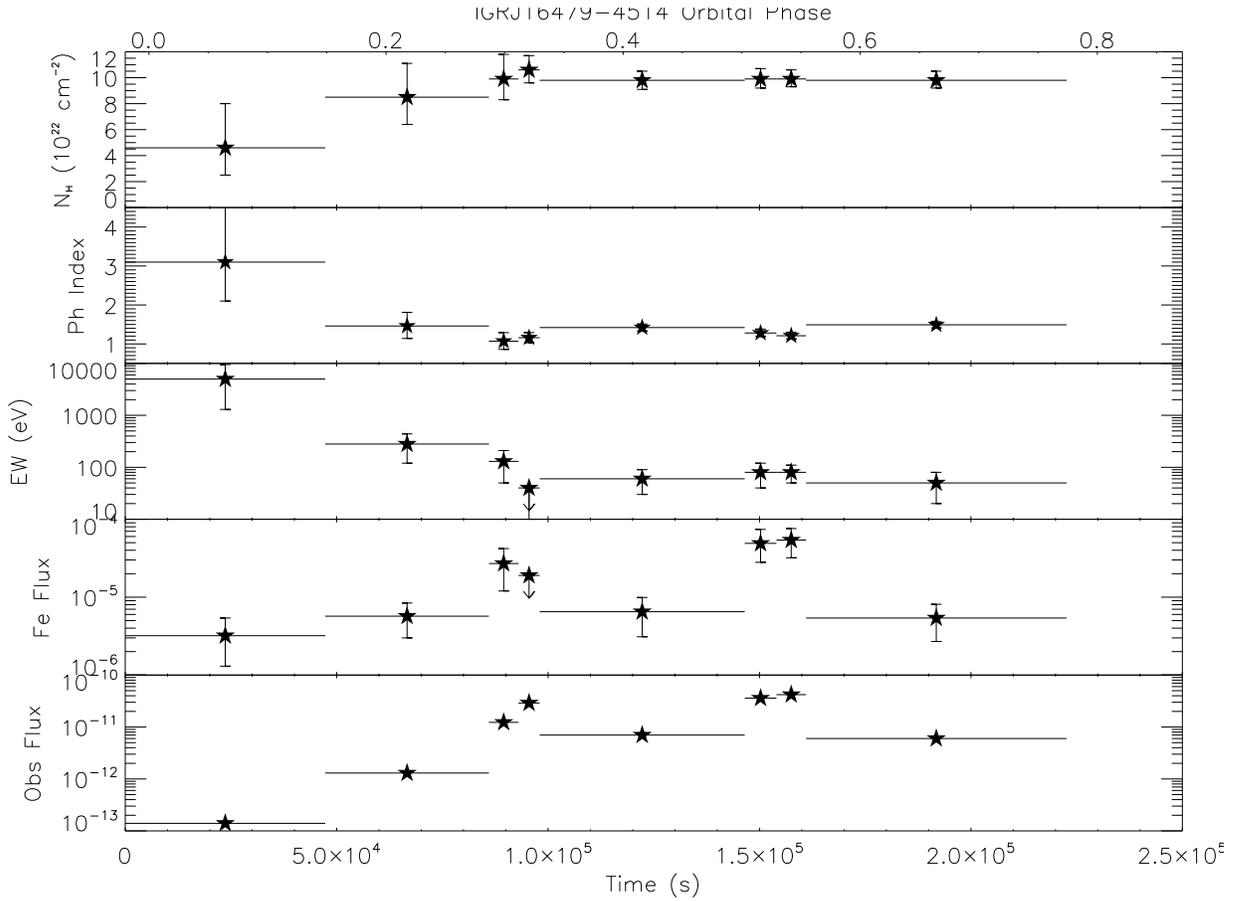}
\end{center}
\vspace{.4truecm}
\caption{\scriptsize $Suzaku$/XIS (1--10 keV)  temporal selected 
spectroscopy of IGR~J16479-4514 (the eight time
intervals are marked in Fig.\ 1). The upper x-axis reports 
the orbital phase, assuming the orbital period of 286792~s (Romano et al. 2009) 
and epoch 54547.05418 MJD (Bozzo et al. 2009).
Spectra have been fitted with an absorbed power law together with a narrow
Gaussian line at 6.4 keV. From top to bottom, the spectral parameters are the following:
absorbing column density, power law photon index, 
equivalent width (EW) of the neutral iron line (in eV)
measured with respect to the power law continuum, 
flux of the iron line (in units of photons cm$^{-2}$ s$^{-1}$),
observed flux (1--10 keV) in units of erg cm$^{-2}$ s$^{-1}$.
}
\label{lsfig:suz_sp}
\end{figure}

\clearpage

The eight spectra could be well described by an absorbed power law together
with a narrow Gaussian line with energy consistent with 6.4 keV, produced by
fluorescence from cold iron in the supergiant wind.
A summary of the spectral  results can be found in Fig. \ref{lsfig:suz_sp}.

Outside the eclipse, the X--ray luminosity is at a level of 10$^{34}$\ergsec, assuming a distance of 2.8 kpc
(intra-flares X--ray emission). 
The power law photon index is harder when the source is brighter, as in other SFXTs and
HMXBs accreting pulsars.

The absorption does not show evidence for variability, within the uncertainties,
except during the eclipse, where it is lower. This is consistent with
the presence of Thomson scattering by electrons in the supergiant wind (Lewis et al. 1992). 

The equivalent width of the Fe~K${_\alpha}$ emission line is around 100 eV in the uneclipsed X--ray emission,
while it is much larger during the eclipse (this is due to the fact that it is measured
with respect to the scattered X--ray emission, the only component visible in this part of the orbit).
The intensity of the iron emission line   
is variable along the orbit and  correlates with the unabsorbed X--ray flux above 7 keV,
as expected (Inoue 1985).

 \section{A mechanism reducing the accretion rate?}

The scattered X--ray emission observed during the X--ray eclipse  
can be used to estimate the density of the supergiant wind 
at the orbital separation,  a (a=2.2$\times10^{12}$~cm assuming a 3.32 days orbital period, a
companion mass, M$_{opt}$, of 35~\msun, and a neutron star of 1.4~\msun).

X--rays observed during the eclipse are about 5\% of the
uneclisped, intra-flare, X--ray intensity. 
The supergiant wind density  can be estimate 
as $n_{\rm w}$ = $0.05 / (a \sigma_{\rm T})$, where  $\sigma_{\rm T}$ is the Thomson cross section.
This results into a wind density at the orbital separation, $\rho$$_{\rm w}(a)$, of 7$\times$10$^{-14}$~g~cm$^{-3}$.
Adopting the mass continuity equation, we can derive the ratio between the wind mass-loss rate, $\dot{M}_{w}$,
and the wind terminal velocity, v$_{\infty}$, as $\dot{M}_{w}/ v_{\infty}$=4~$\pi$~$a~(a-R_{opt}$)~$\rho_{\rm w}(a)$,
assuming a spherical outflowing wind and a velocity law for the wind velocity with $\beta$=1 (Castor et al. 1975).

The resulting ratio is $\dot{M}_{w}/ v_{\infty}$=7$\times$10$^{-17}$~M$_{\odot}$/km.
If we assume reasonable wind terminal velocities in the interval 500--3000~km~s$^{-1}$, 
the mass loss rate is $\dot{M}_{w}$=1--7$\times$10$^{-6}$~M$_{\odot}$/yr.

Assuming Bondi-Hoyle accretion from the supergiant wind, the accretion rate onto the neutron star 
can be estimated from the relation
$\dot{M}_{\rm acc}$=$(\pi R_{acc}^{2} / 4\pi a^{2})$$\cdot$$\dot{M}_{w}$, where R$_{acc}$ is the accretion radius. 
This accretion rate implies an X--ray luminosity L$_{\rm X}$=3--15$\times$10$^{36}$~erg~s$^{-1}$,
which is at least two orders of magnitude larger than observed (L$_{\rm X}$$\sim$10$^{34}$~erg~s$^{-1}$).

Given that the wind density we have derived here is too high for the observed accretion luminosity, 
the  low X--ray luminosity in \src\ is unlikely to be produced by direct wind accretion, implying
the presence of a mechanism able to lower the accretion rate and damp
the X--ray variability implied by the structured supergiant wind. 
This is in agreement with the results recently obtained by Oskinova et al. (2012) in HMXBs.

A possibility is that the   magnetospheric surface mediates the accretion, acting 
to reduce the accretion rate onto the 
neutron star, as already suggested by several authors to explain the peculiar X--ray behavior of the
members of the SFXTs class 
(Bozzo et al. 2008b, Ducci et al. 2010, Shakura et al. 2012, Postnov et al. 2013, these proceedings).

\acknowledgments
This work was supported in Italy by ASI-INAF contracts I/033/10/0 and I/009/10/0, and by 
the grant from PRIN-INAF 2009, ``The transient X--ray sky: 
new classes of X--ray binaries containing neutron stars''
(PI: L. Sidoli). AB received funding from NASA grant 11-ADAP11-0227.


\begin{scriptsize}

\end{scriptsize}


\begin{thebibliography}{99}
\bibitem{} Bozzo, E., Stella, L., Israel, G., et al., 2008a, \emph{MNRAS},  {\bf 391}, L108 
\bibitem{} Bozzo, E., Falanga, M., Stella, L., 2008b, \emph{ApJ}, {\bf 683}, 1031 
\bibitem{} Bozzo, E., Giunta, A., Stella, L., 2009, \emph{A\&A}, {\bf 502}, 21 
\bibitem{} Castor, J.I., Abbott, D.C., Klein, R.I., 1975, \emph{ApJ}, {\bf 195}, 157
\bibitem{} Ducci, L., Sidoli, L., Paizis, A., 2010,  \emph{MNRAS}, {\bf 408}, 1540  
\bibitem{} Inoue, H., 1985,  \emph{Space Science Reviews}, {\bf 40}, 317 
\bibitem{} Jain, C., Paul, B., Dutta, A., \emph{MNRAS}, {\bf 397}, L11
\bibitem{} Koyama, K., et al., 2007, \emph{PASJ}, {\bf 59}, 23
\bibitem{} Lewis, W., et al., 1992, \emph{ApJ}, {\bf 389}, 665 
\bibitem{} Molkov, S., et al., 2003, \emph{The Astronomer's Telegram},  {\bf 176} 
\bibitem{} Negueruela, I., Smith, D.M., Reig, P., et al. 2006, in \emph{ESA Spec. Pub.}, ed. A.Wilson, Vol. {\bf 604}, 165-170 
\bibitem{} Nespoli, E., Fabregat J., Mennickent, R.E., 2008, \emph{A\&A},  {\bf 486}, 911 
\bibitem{} Oskinova, L. M., Feldmeier, A., and Kretschmar, P., 2012, {\em MNRAS}, {\bf 421}, 2820
\bibitem{} Postnov, K.,  et al.,  2013, these proceedings (arXiv:1212.2841)
\bibitem{} Ratti, E. M., et al., 2010, \emph{MNRAS},  {\bf 408}, 1866
\bibitem{} Romano, P., Sidoli, L., Cusumano, G. et al., 2009, \emph{MNRAS},  {\bf 399}, 2021 
\bibitem{} Sguera, V., Barlow, E.J., Bird, A.J., et al. 2005, \emph{A\&A},  {\bf  444}, 221 
\bibitem{} Sguera, V., Bazzano, A., Bird, A. J., et al. 2006, \emph{ApJ},  {\bf  646}, 452 
\bibitem{} Shakura, N., Postnov, K., Kochetkova, A., Hjalmarsdotter, L., 2012, \emph{MNRAS }, {\bf  420}, 216  
\bibitem{} Sidoli, L.; Romano, P.; Mangano, V., et a., 2008, \emph{ApJ}, {\bf 687}, 1230 
\bibitem{} Sidoli, L., Esposito, P., Sguera, V., et al., 2013, \emph{MNRAS}, in press, \emph{arXiv.1212.0723}

\end{thebibliography}
\end{document}